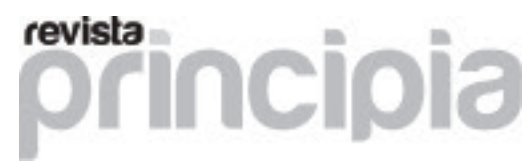

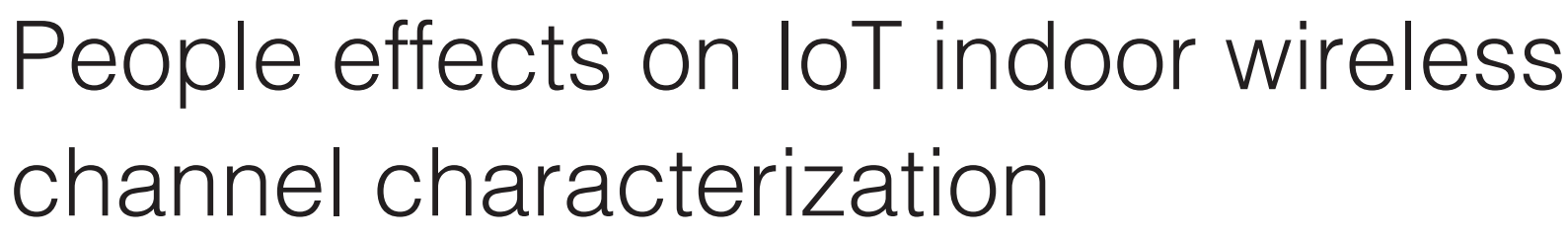

# People effects on IoT indoor wireless channel characterization

Millena Michely de Medeiros Campos [1], Mateus de Oliveira e Mattos [2], Rafael da Silva Macedo [3], Alvaro Augusto Machado de Medeiros [3], Wellerson Viana de Oliveira [4], Vicente Angelo de Sousa Jr [4]

[1] millena.campos@sidia.com, Sidia Institute of Science and Technology, Manaus, AM, Brazil. [2] mateus.mattos@engenharia.ufjf.br, Centre for Research and Development in Telecommunications (CPqD), Campinas, SP, Brazil.
[3] {rafael.macedo2016, alvaro}@engenharia.ufjf.br, Federal University of Juiz de Fora, Juiz de Fora, MG, Brazil.
[4] {wellerson, vicente.sousa}@ufrn.edu.br, Federal University of Rio Grande do Norte (UFRN), Natal, RN, Brazil.

## ABSTRACT

Wireless communication under 1 GHz is suitable for Internet of Things (IoT) applications due to larger coverage capability with less power consumption. Bearing in mind that people and elements contained in the environment can cause variations in the channel, this paper aims to evaluate the effect of the presence of people on a 900-MHz indoor narrowband wireless channel, as we characterize the small-scale phenomena. With the increase in the number of people, a greater variation in the communication channel was noticed, which is reflected in the parameters of the probability distributions used in the characterization of the random part of the signal. In addition, second-order statistics were used to analyze the data and an adherence test was applied to confirm the behavior of the signal in relation to the distributions.

## RESUMO

*A comunicação sem fio usando frequências abaixo de 1 GHz é adequada para aplicativos de Internet das Coisas (IoT) devido à maior capacidade de cobertura com menor consumo de energia. Tendo em vista que pessoas e elementos contidos no ambiente podem causar variações no canal, este artigo tem como objetivo avaliar o efeito da presença de pessoas em um canal sem fio interno de banda estreita de 900 MHz, conforme caracterizamos os fenômenos de pequena escala .Com o aumento da quantidade de pessoas, percebeu-se uma maior variação no canal de comunicação que é refletida nos parâmetros das distribuições de probabilidades utilizadas na caracterização da parte aleatória do sinal. Além disso, estatísticas de segunda ordem foram utilizadas na análise dos dados e um teste de aderência foi aplicado para confirmar o comportamento do sinal em relação às distribuições.*

## 1 Introduction

Future wireless communication systems are required to provide access to a great number of terminals. This occurs not only because of the increasing number of computers and cellphones but also to the concept of connection of as many objects as possible to the Internet, the so-called Internet of Things (IoT). As a result, new proprietary technologies, e.g., LoRa (SEMTECH, 2020) and Sigfox (SIGFOX, 2020), have been developed to meet this increasing demand. Standardization organizations such as 3GPP (3rd Generation Partnership Project), IEEE and ITU (International Telecommunications Union) have also worked to define standards for Machine-to-Machine (M2M) communications, such as NB-IoT (Narrowband IoT) from 3GPP (3GPP, 2020), Wi-Fi HaLow 802.11ah from IEEE (WI-FI ALLIANCE, 2018) and the use case 5G massive Machine Type Communications from ITU (ITU-R, 2015). Many of these technologies use narrowband channels of the Industrial, Scientific and Medical band (ISM), such as the 900 MHz frequency band.

The design and implementation of these technologies require the study and characterization of propagation in environments where they will operate. However, outdoor propagation or free space models are not always valid for indoor environments, since they impose a higher variability to the wireless channel even with smaller distances between transmitter and receiver (RAPPAPORT, 2002). Besides, the characteristics of the environment itself, such as walls, furniture, and the flow of people also affect the propagation.

Several works addressed the indoor propagation characterization. Some evaluate the propagation in the range of 2.4 GHz (WALKER; ZEPERNICK; WYSOCKI, 1998; JANSSEN; PRASAD, 1992), others in the UHF range below 1 GHz (MAYER; WRULICH; CABAN, 2006; RAO; BALACHANDER; TIWARI, 2012; SCZYSLO; DORTMUND; ROLFES, 2012). Some studies assess the effects of people's presence and movement in indoor propagation at 2 GHz, observing a high channel variability proportional to the increasing numbers of people in the environment (KARA; BERTONI, 2006; WANG; LU; ZHU, 2013).

Several recent studies have proposed human gesture recognition systems (SHAHZAD; ZHANG, 2018; AHMED et al., 2019; LEE et al., 2019; REGANI et al., 2020), using the commercial Wi-Fi signal. It is possible to recognize the different influences produced by people and their movements in the communication channel, but it requires a large database that enables the system to learn how to recognize the commands. However, there is a lack of studies evaluating the effect on frequencies below 1 GHz, especially in IoT scenarios with the high occupation. This is the main scope of this contribution which aims to evaluate the effect of the presence of people on the indoor wireless channels through 900-MHz narrowband measurements. The measurements were carried out at the refectory of the Federal University of Juiz de Fora (UFJF), simulating a real IoT system operation in an environment with a high density of people.

The paper is organized as follows. Section 2 describes our low-cost experimental setup. Both data capturing and post-processing are described in Section 3. Section 4 shows and discusses the results, and Section 5 presents the final comments.

## 2 Measuring System

Our experiment consists of a narrowband channel setup based on a RF generator as a transmitter. On the other side, a receiver based on Software-Defined Radio (SDR) platform tuned to this frequency allows interfacing with a computer for data collection and post-processing.

The Agilent RF generator N9310A (KEYSIGHT, 2014) set for the transmission of a continuous wave (CW) at 900 MHz was used with a power of 0 dBm. A vertically polarized microstrip dipole antenna attached to a PVC rod was used to assure the antenna remained 1.60 m above the ground. On the receiver side, a SDR platform, composed of an Universal Software Radio Peripheral (USRP) N210 (ETTUS RESEARCH, 2020a) and a personal computer, replaced components commonly used in dedicated hardware. Thus, the computer is used to control the capture of measurements using the RF daughterboard WBX 50-2200 MHz Rx/Tx (ETTUS RESEARCH, 2020b), at 10 MHz sample rate, attached in the USRP. Transmitting and receiving antennas have the same specification in order to better capture the multipath effects due to their omnidirectional radiation pattern. The height and placement of the antennas were chosen to simulate IoT applications, such as wearables or sensors. A notebook computer with an Intel processor i5 6200U, using the Linux Manjaro distribution (LINUX, 2020), controls the USRP N210 and the measurement process. We developed a script for data measurement and storage. Figure 1 presents the transmitter and receiver setups.

We validated the USRP receptor using the Agilent N9310A RF generator. We connected both via RF cable with a 10 dB attenuator at the 900 MHz frequency.

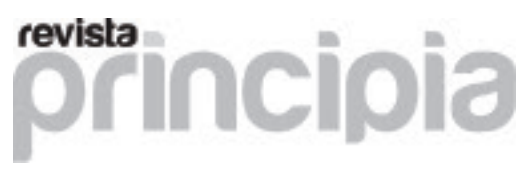

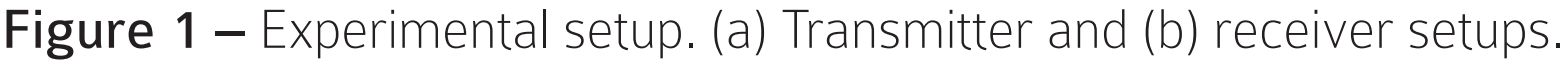

**Figure 1 –** Experimental setup. (a) Transmitter and (b) receiver setups.

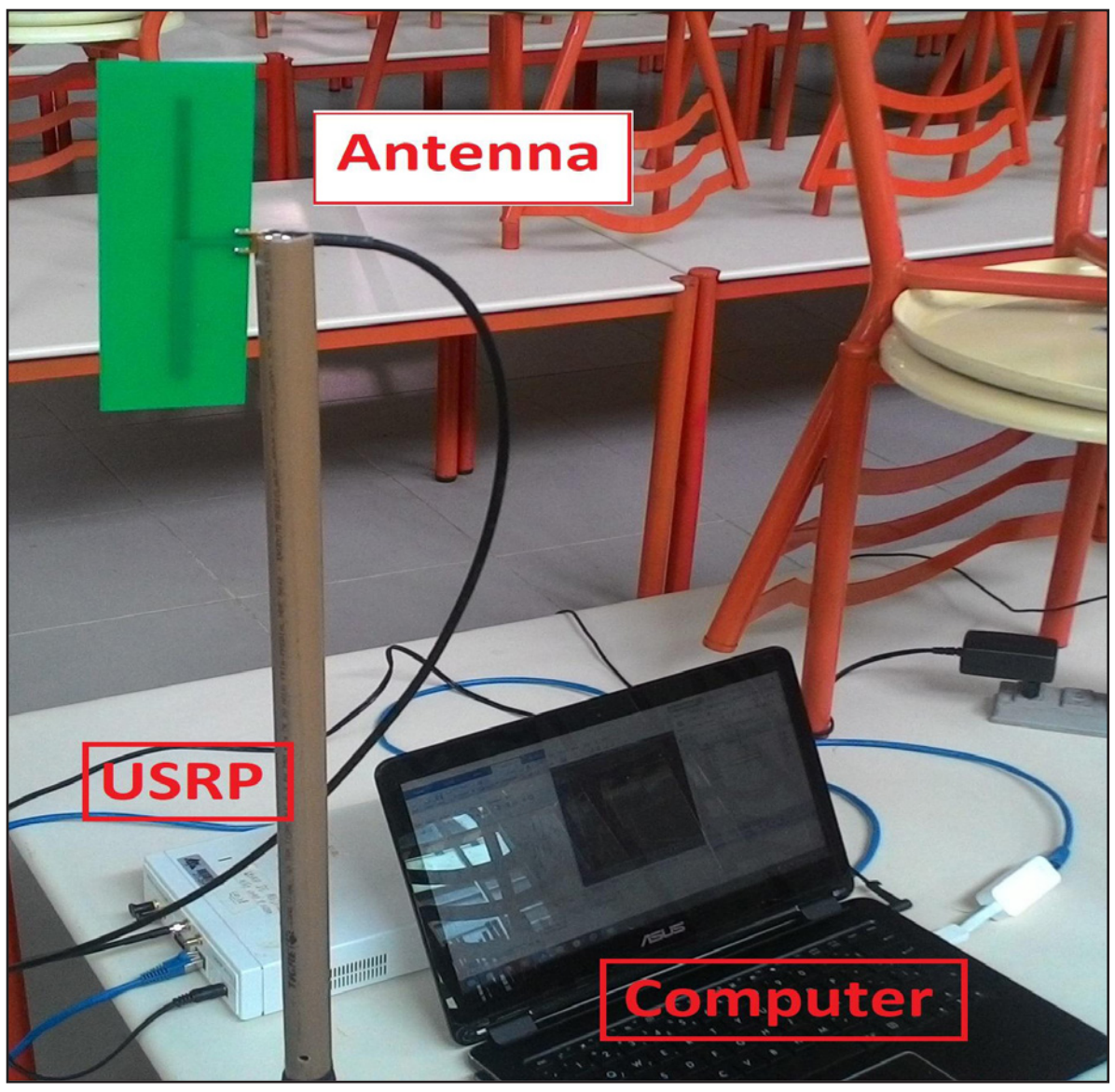

(a) (b)

*Source:* Authors.

As presented in Figure 2, there is linearity in signal reception in the range of -80 to 5 dBm of transmitting power variation.

**Figure 2 –** USRP N210 calibration graph

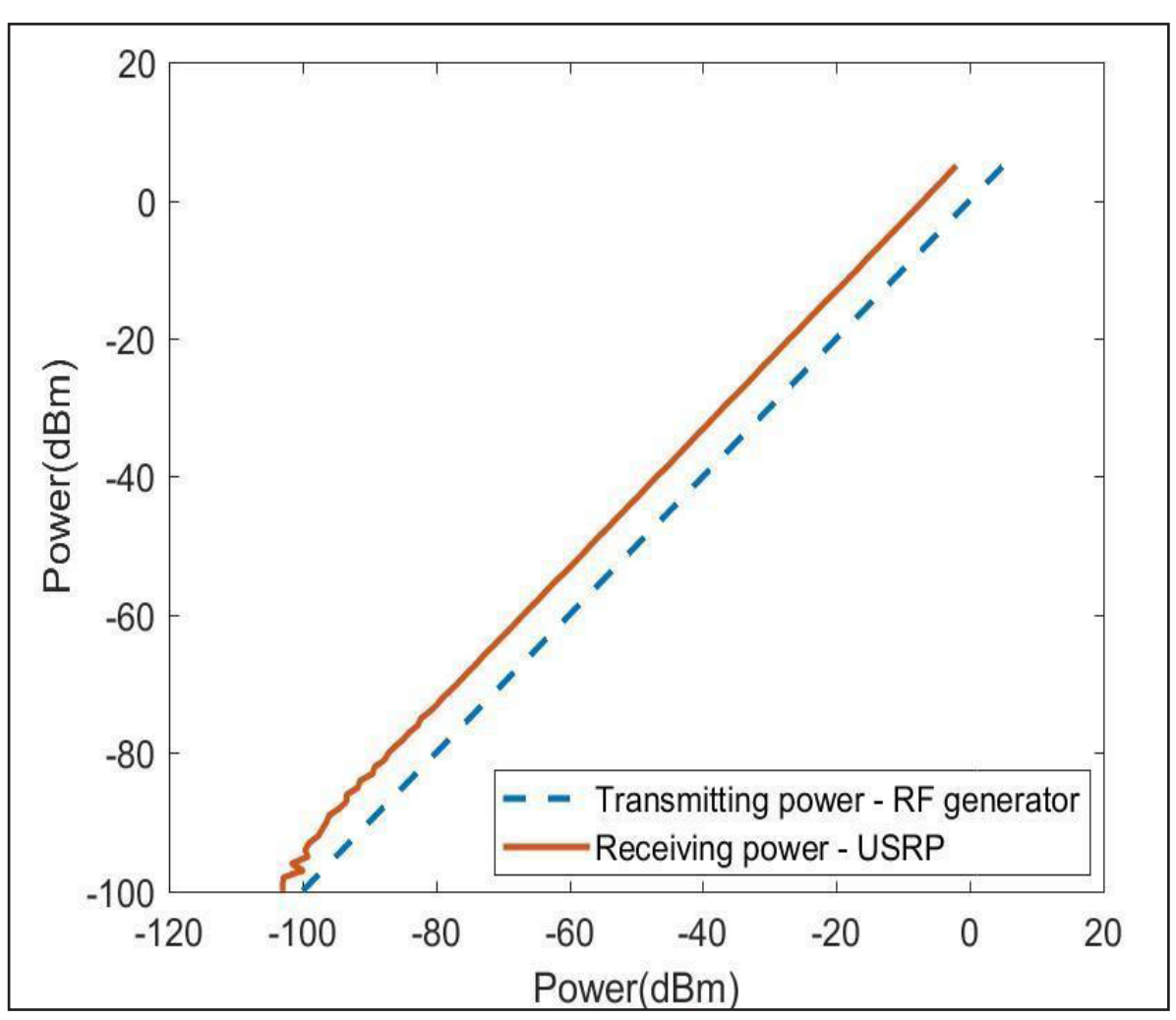

*Source:* Experimental data.

## 3 Data measurement and processing methods

Three measurement campaigns were carried out on three different days in February 2018, at the refectory during its opening hours (11:00 a.m. to 2:00 p.m.). The refectory was operating at half of its capacity due to the undergraduate vacation period, with a daily flow of approximately 1200 people during the measurements. The number of people was counted each 30-minute interval, resulting in categories with different average numbers of people.

Figure 3 presents the refectory floor plan, indicating which tables were not being used during the measurement period (crosshatched tables), as well as a photography of the site with people. For all measurements, the transmitter and receiver were placed 35 m apart from each other at the points A and B, respectively. Antenna locations were configured to provide line-of-sight (LOS) propagation between transmitter and receiver. The sampling time was 0.5 s. Thus, we analyzed two scenarios: (i) Scenario 1: empty restaurant; and (ii) Scenario 2: crowded restaurant.

We use a sliding window filter to remove large-scale variations from the measured signal, where the average value of the envelope signal is calculated for a group of adjacent samples of each measured sample (YACOUB, 1993). After that, the normalized envelope is obtained by dividing the measured envelope signal by envelope RMS value. The resultant signal denotes the small-scale normalized envelope signal, which we use to analyze the variations of small-scale fading (COTTON; SCANNLON, 2007).

**Figure 3 –** Measurement location: floor plan and photo.

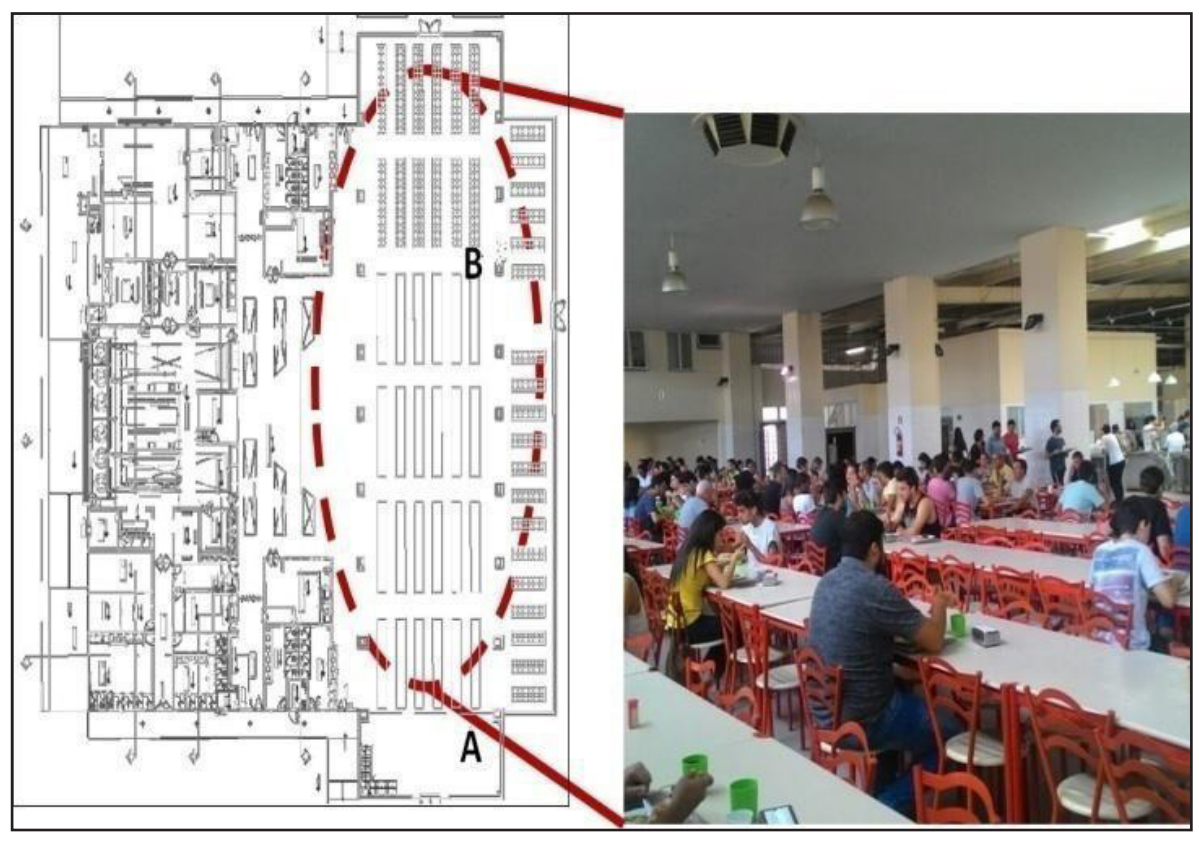

*Source:* Authors.

We analyzed the empirical cumulative distribution function (CDF) of the small-scale fading samples with three well-known fading distributions, as described as follows (YACOUB, 1993; KATTENBACH; ENGLERT, 1998; COTTON; SCALON, 2007):

1. Rice: the Rice distribution is used for small-scale fading modeling of environments that have LOS propagation. The CDF of the normalized envelope $\rho$ is given by

$$F_p(\rho)=1-Q_1\left(\sqrt{2k},\sqrt{2(k+1)}\rho\right), \qquad (1)$$

where $k$ is the Rice factor and $Q_1(\cdot)$ the Marcum-Q function. The classic Rayleigh distribution, which assumes the absence of a deterministic LOS component, can be considered a particular case of Rice distribution when $k = 0$.

Nakagami: The Nakagami distribution models channel present multipath clusters, i.e. groups of multipath components with similar delays and phase differences. Its envelope CDF in the normalized form is

$$F_p(\rho)=1-\frac{\Gamma(m,m\rho^2)}{\Gamma(m)}, \qquad (2)$$

where $\Gamma(\cdot,\cdot)$ is the incomplete Gamma function and $m$ the Nakagami parameter. When $m = 1$, Nakagami and Rayleigh distributions are equivalent.

3. Weibull: The Weibull is used to model channels with multipath fading with non-linearities. The CDF of the normalized envelope is

$$F_p(\rho)=1-e^{-\rho\alpha} \qquad (3)$$

where $\alpha$ is related to the non-linearity of the propagation medium. The linear fading system can be obtained with $\alpha = 2$, and the distribution equals to the Rayleigh distribution.

For the statistical analysis of the data, we applied the non-parametric Kolmogorov-Smirnov Statistical Test (KS Test). It quantifies the distance between the empirical CDF of the measurements and the reference CDF, informing if the hypothesis that the empirical CDF fits the analytic CDF is true or false.

There is a considerable number of parameter estimators for the evaluated models. Several works use the Maximum Likelihood Estimator (MLE) due to the feasibility of its estimation, even for large data sets (DUMOUCHEL, 1971; NOLAN, 2001). Additionally, there is a fast algorithm to compute the MLE, and it achieves the Cramer-Rao bound (BODENSCHATZ, 1999). For these reasons, the MLE is used as the estimator for studies herein presented.

Generally speaking, the MLE method estimates the parameters $\theta$ which specifies a probability function $f(x_i \vee \theta)$ of a random variable $X$ (MYUNG, 2003). The estimation is based on the independent and identically distributed (i.i.d.) samples $x_i$ (observations) from the distribution, and a log-likelihood function $l(\theta)$, which is given by

$$l(\theta)=\sum_{i=1}^{N} \square logf(x_i \vee \theta) \qquad (4)$$

Thus, the MLE chooses the model parameters $\theta$ that maximize the likelihood function, yielding the most likely parameters to generate the observed data (MIURA, 2011).

The parameters $k$, $m$ or $\alpha$ of each normalized CDF were estimated by MLE and the error of the KS Test is calculated as $D_n = sup_x \vee F_n(x) - F(x) \vee$, where $F(x)$ is the target CDF and $F_n$ the empirical CDF.

## 4 Results and Discussions

For illustration purposes, Figure 4 depicts an example of the received power signal in a time interval of 450s (900 samples) for measurements with and without people. There is a difference of almost 10 dB on the average power of each scenario even with LOS reception. We also observe almost insignificant variation on samples obtained for Scenario 1. Thus, as described in Section 3, we extract small-scale samples from Scenario 2 samples and analyze them as follows.

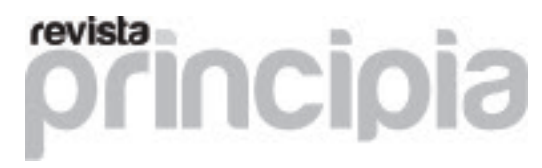

**Figure 4 –** Samples of received power over 450 s for the two tested scenarios.

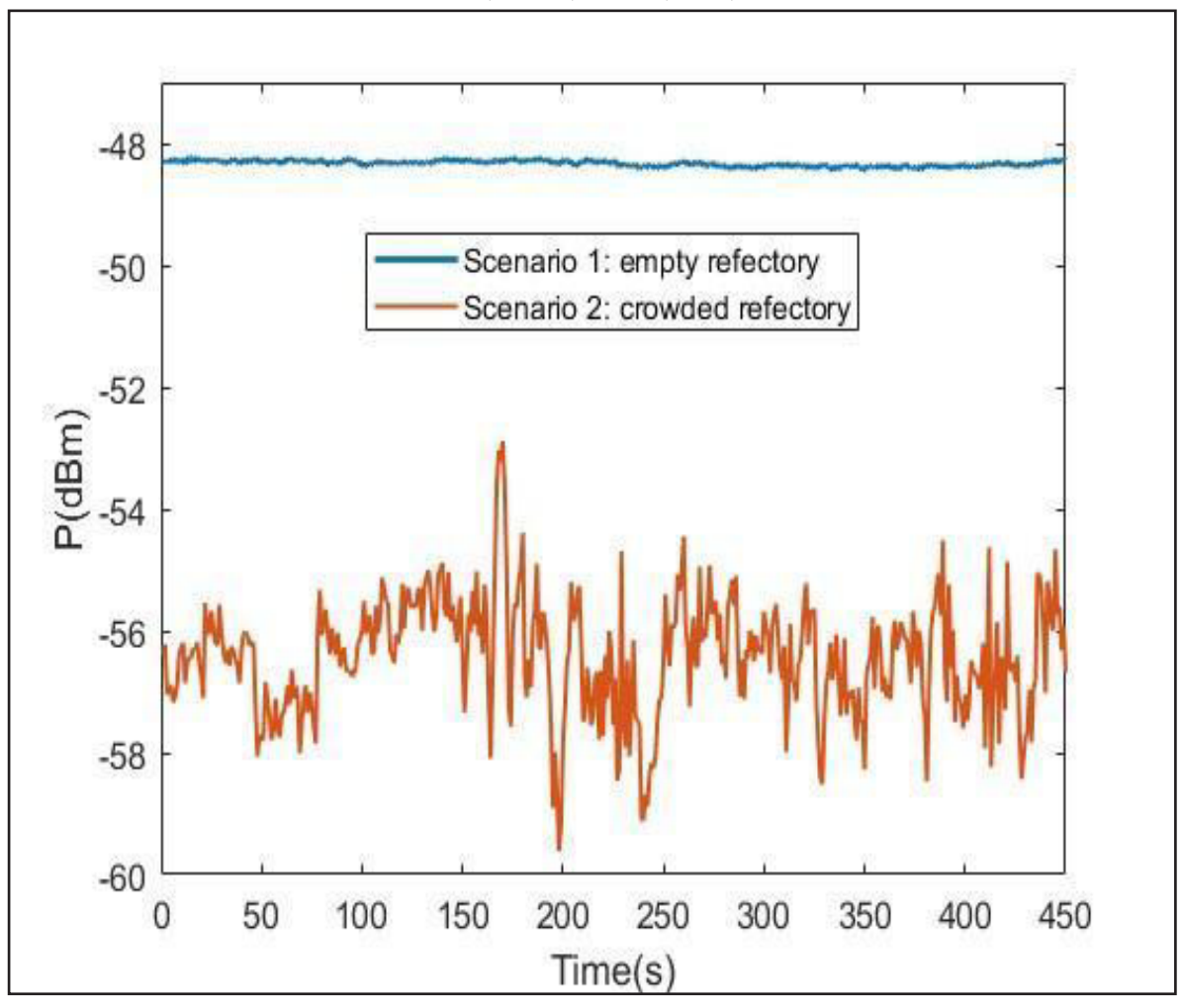

*Source:* Experimental data.

## 4.1 Analyzing the complete dataset

In order to analyze the measured environment regardless of the number of people, we present the small-scale fading characterization for the whole collected data in the refectory to Scenario 2. Figure 5 shows the empirical CDF as well as theoretical CDFs with the estimated parameters by MLE. We notice the difficulty to fit measured data to theoretical functions due mainly to the diversity of the measurement conditions. This result indicates that the measured data should be somehow categorized for a more accurate analysis. Thus, we analyze the measured data in function of the number of people at the restaurant.

**Figure 5 –** Empirical and theoretical CDFs (parameters by the MLE).

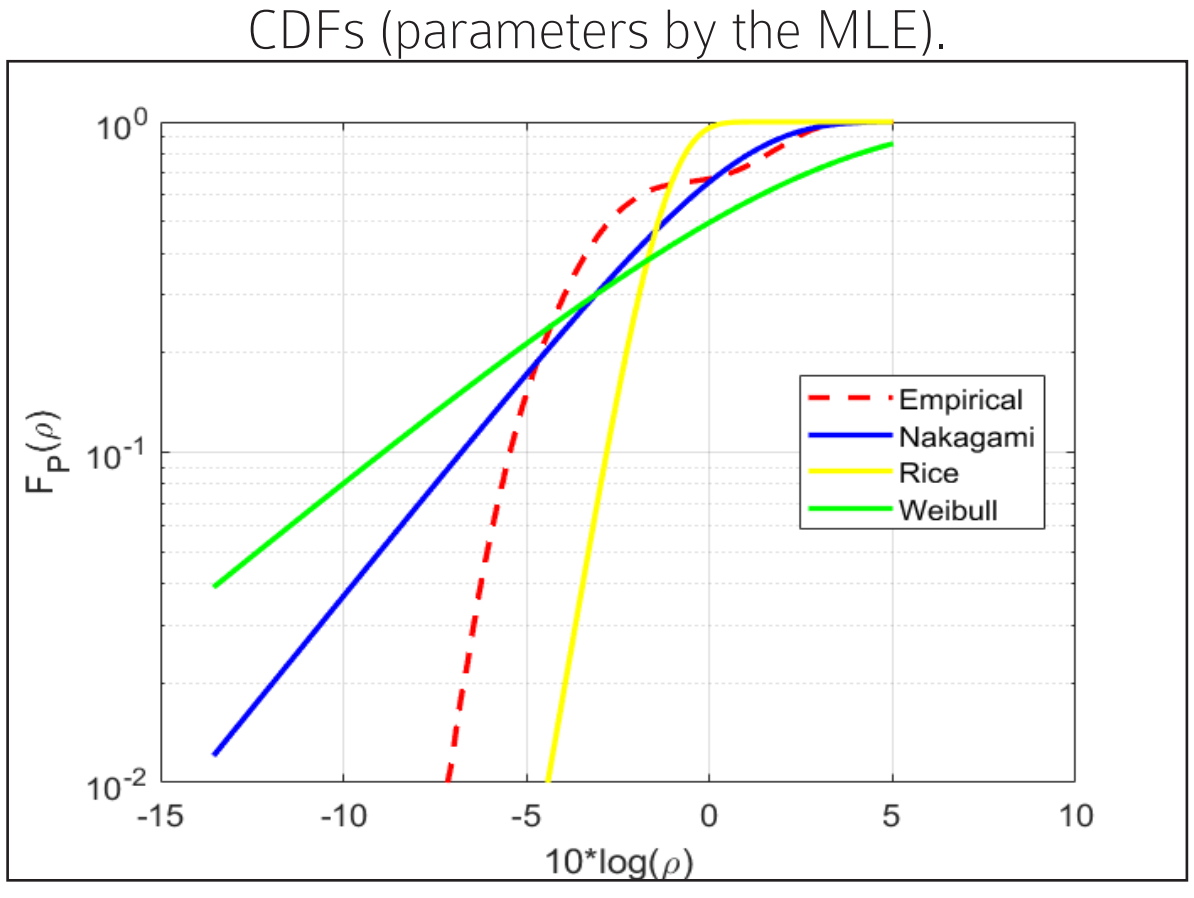

*Source:* Experimental data.

## 4.2 Analyzing the influence of people

The influence of the different quantity of people on three measurement campaigns of Scenario 2 (crowded restaurant) can be observed in Table 1. It presents the estimated parameters for all tested distributions for an average number of people. The results indicate a variation of the distribution parameters as the occupation of the restaurant changes. Table 1 shows an inverse dependence between the number of people and the value of the distribution parameters, indicating higher channel variability with the increase of the number of people.

**Table 1 –** Parameters obtained for distributions.

| Campaign | Number of people | Parameters | | |
|---|---|---|---|---|
| | | $k$ | $m$ | $\alpha$ |
| 1 | 56 | 61.835 | 30.584 | 11.569 |
| | 116 | 62.587 | 31.321 | 10.800 |
| | 171 | 30.621 | 15.645 | 8.139 |
| | 212 | 21.758 | 11.366 | 6.462 |
| 2 | 100 | 49.896 | 24.631 | 9.447 |
| | 133 | 43.137 | 21.813 | 8.973 |
| | 233 | 27.281 | 13.598 | 7.621 |
| | 272 | 27.762 | 14.136 | 7.780 |
| 3 | 111 | 93.406 | 45.971 | 11.482 |
| | 133 | 69.600 | 33.195 | 11.733 |
| | 138 | 56.080 | 27.921 | 9.867 |
| | 264 | 49.175 | 24.542 | 10.249 |

*Source:* Authors data.

Figure 6 presents the KS test error of the estimated parameters for Campaign 1, i.e. the first day of measurements. The KS test indicates the Nakagami as the distribution which fits more properly with the empirical data. We also notice the smaller errors for a higher number of people due to a higher variation of power level, which yields different values of $\rho$. The KS test error is meaningfully sensible to distributions that exhibit heavy tails, especially when the collected data does not entirely span the domain of the distribution function. The other campaigns present the same qualitative tendency.

**Figure 6 –** KS error of each estimated distribution for Campaign 1.

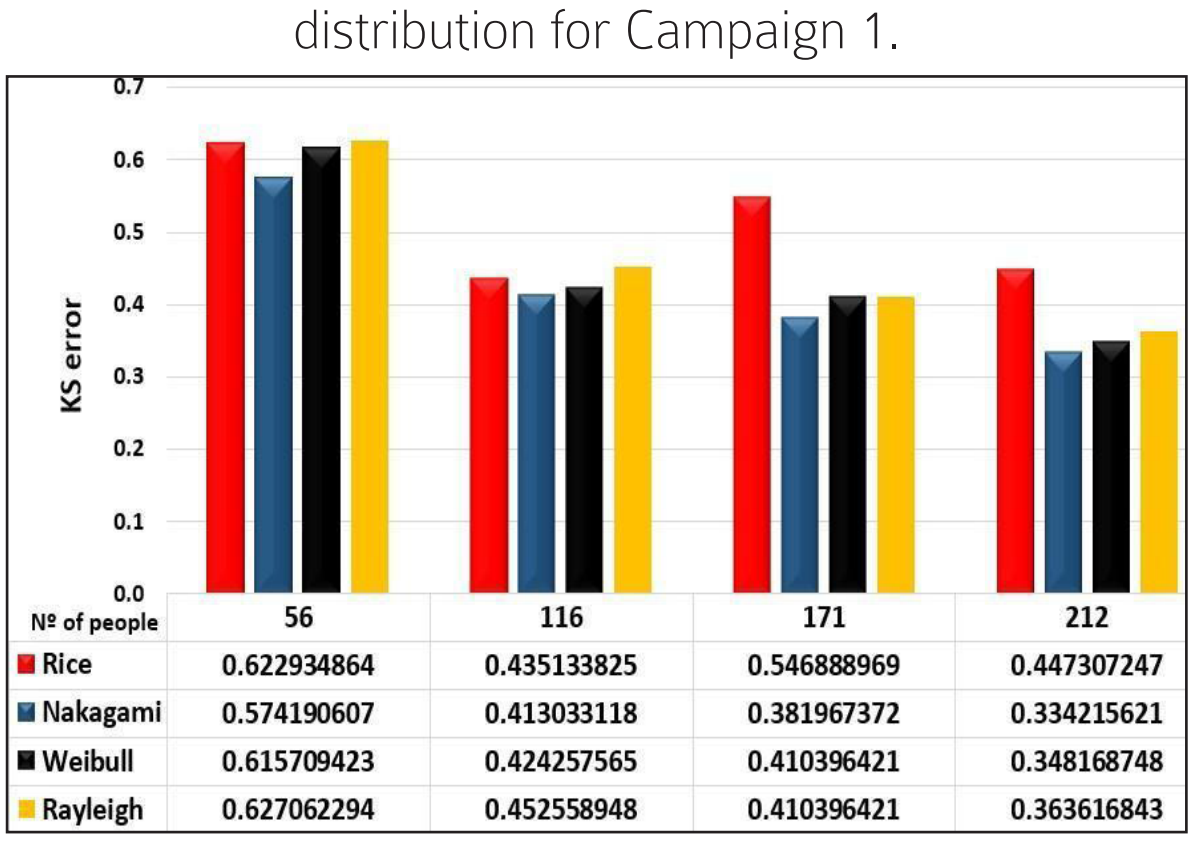

*Source:* Authors data.

As we categorize the data in intervals of 50 people, we also observe the best fit of the Nakagami distribution to measured data. The values of the $m$ parameter for these intervals are shown in Table 2. For illustration purposes, Figure 7 depicts the comparison of theoretical distributions with measured data for the interval of 250 to 300 people.

**Figure 7 –** Empirical and theoretical CDFs for the interval of 250 to 300 people (parameters calculated using the MLE).

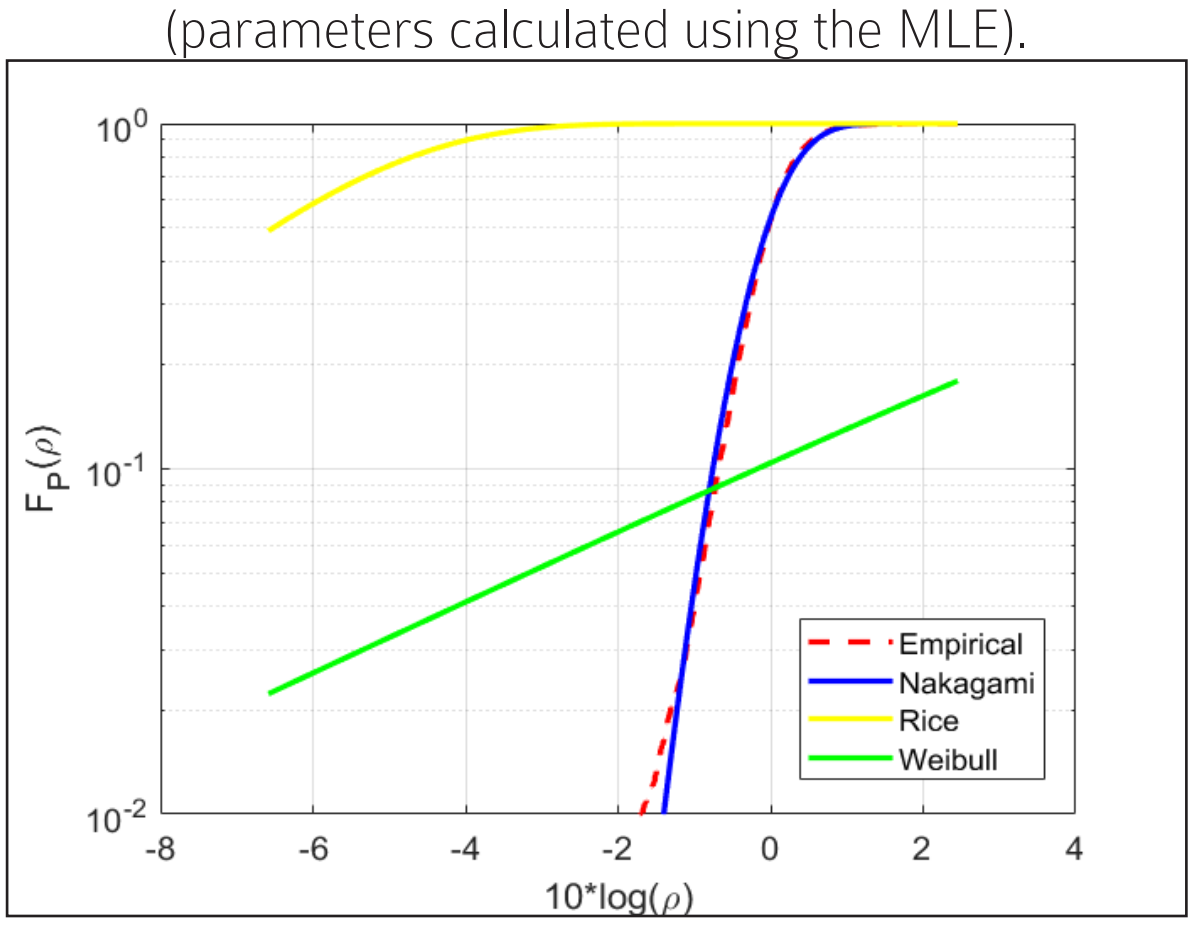

*Source:* Authors data.

**Table 2 –** Estimated $f_d$ according to quantity of people.

| Quantity of people | $m$ | Estimated mean $f_d$ |
|---|---|---|
| 50 - 100 | 24.6314 | 0.1106 |
| 100 - 150 | 21.8132 | 0.1163 |
| 150 - 200 | 15.6451 | 0.1388 |
| 200 - 250 | 11.3655 | 0.1486 |

*Source:* Authors data.

## 4.3 Second-order analysis

A time-varying small-scale fading could be experienced even when transceivers are stationary, mainly due to the movement of surrounding scattering objects (RAPPAPORT, 2002). An important second-order statistic that characterizes the time-varying mobile channel is the Level Crossing Rate (LCR). The LCR corresponds to the rate at which the received signal envelope intensity crosses a certain threshold in the positive direction (YACOUB, 1993; COTTON; SCALLON, 2007). The LCR is closely related to the maximum Doppler shift, which is a classical parameter to model the time-varying channel phenomenon. Thus, LCR closed-form expressions can be used to estimate the maximum Doppler shift $f_d$ of fading sample data.

Based on the results of Section 4.2, we selected the Nakagami as the fading distribution that fits properly with the distribution of the measured data. Therefore, by analyzing the measured samples, we can empirically estimate the LCR, namely $LCR_{Empirical}$. We can find the estimated maximum Doppler shift $f_d$ by means of

$$f_d = \frac{LCR_{Empirical}}{LCR_{Theoretical}} \quad (5)$$

and

$$LCR_{Theoretical} = \sqrt{2\pi}\,\frac{m^{m-0.5}}{\Gamma(m)}\rho^{2m-1}\exp\exp\left(-m\rho^2\right) \quad (6)$$

where $LCR_{Theoretical}$ is given by normalized expressions ($f_d$=1) of the LCR for the Nakagami distribution (YACOUB; BAUTISTA; GUEDES, 1999). $\Gamma(m)$ is the Gamma function, $m$ is the Nakagami distribution parameter and $\rho$ is the normalized envelope.

Figure 8 presents the maximum Doppler shift estimated for some signal level thresholds using LCR. As expected, the estimation provides stable results for the specific region of signal levels where the threshold crossing is intense. Such a signal threshold region is close to zero, where we are analyzing the variation of the signal around the mean. We divided the measurement data into classes of 50 people. As the quantity of people increases, the value of $f_d$ becomes more stable, even for lower thresholds. It occurs because the envelope variation intensifies as the number of people increases, as discussed in Section 4.2, and consequently we have more level crossings in lower thresholds.

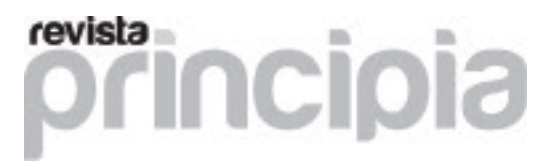

Although the parameter of the distribution is influenced by the increase in the number of people, the same may not occur to the maximum Doppler shift $f_d$. Table 2 shows the average values of the estimated $f_d$ for the signal level threshold for levels between -2 and 2 dB. It can be noticed that, regardless of the number of people, the $f_d$ variation is not very significant, since the LCR curves tend to be similar for high values of $m$, as those presented in Table 2, and for signal levels near to 0 dB.

**Figure 8 –** Estimated for some signal level thresholds.

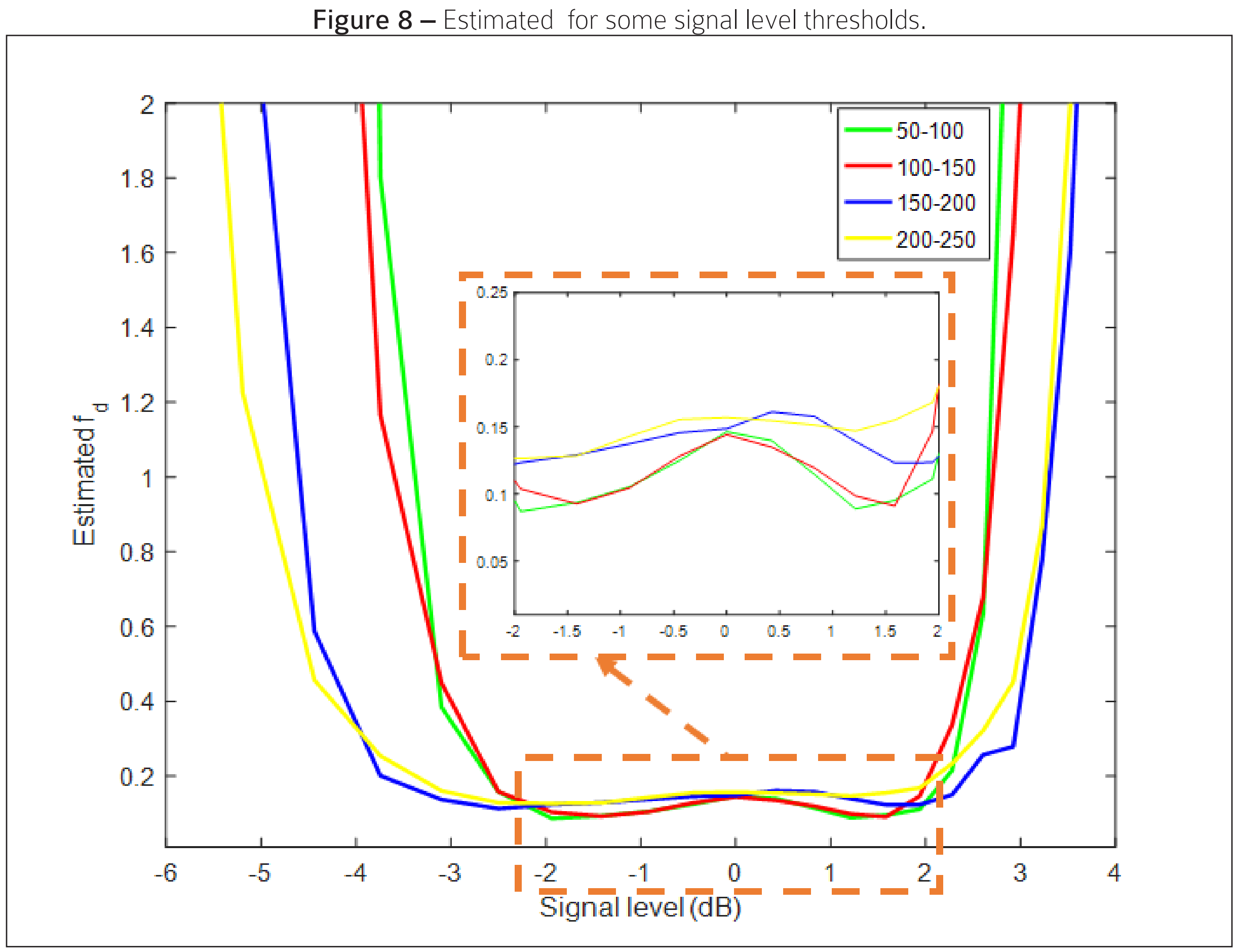

*Source:* Experimental data.

## 5 Conclusions

With the advent of the Internet of Things, sub-1 GHz wireless systems have been developed to provide such type of service. This paper presented an analysis of the effects of the presence of people on the reception of a RF narrowband 900 MHz signal. We realized measurements using low-cost SDR equipment in an environment with up to a few hundreds of people.

Analyzing measurement data collected over three days, we observed different channel variability when comparing the environment with and without people. We calculated the parameters of small-scale fading distributions that fit properly to measured data, and we noticed that the number of people in the environment significantly influences such parameters. We also noticed that the Nakagami distribution is the most suitable to describe the small-scale variations of our propagation scenario.

Finally, we use LCR to estimate the maximum Doppler shift based on the measured data and the LCR theoretical formulation for the best-fitted distribution. Despite affecting the $m$ parameter of the Nakagami distribution considerably, the number of people did not influence the estimated Doppler shift, if one considers levels in the proximity of the envelope's mean value. In our future work, we intend to evaluate the influence of antenna height and positioning, as well as indoor and outdoor combined scenarios.

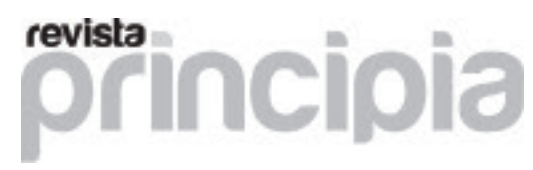